\documentclass[12pt,preprint]{aastex}            

\newcommand{\ergcm}{ergs~cm$^{-2}$~s$^{-1}~$}

\def\erg/cm2sec{ergs~cm$^{-2}$~s$^{-1}$}   
\def\ergcm2{ergs~cm$^{-2}$}   
\def\cm2{cm$^2$} 
\def\la{\hbox{\rlap{\raise.3ex\hbox{$<$}}\lower.8ex\hbox{$\sim$}\ }}
\def\ga{\hbox{\rlap{\raise.3ex\hbox{$>$}}\lower.8ex\hbox{$\sim$}\ }}

\received{}
\accepted{}
\begin{document}

\title{Binary and Long-Term (Triple?) Modulations of 4U 1820$-$30 in NGC
6624} 

\author{Y. Chou and J. E. Grindlay}
\affil{Harvard-Smithsonian Center for Astrophysics, \\
60 Garden Street, Cambridge, MA 02138\\
yichou@cfa.harvard.edu}

\begin{abstract}

We present timing analysis results for Rossi x-ray Timing Explorer
(RXTE) observations of x-ray binary source 4U 1820$-$30 located in the
globular cluster NGC 6624. The light curves of observations made
between October 1996 and September 1997 show that the maximum of the
685s binary period modulation folded by the linear ephemeris from
previous observations has phase shift of $-$0.20 $\pm$ 0.06.  Combined
with historical results (1976-1997), the binary period derivative is
measured to be $\dot P / P =(-3.47 \pm 1.48) \times 10^{-8}$
yr$^{-1}$.  The previous known ($\sim$176 d) long-term modulation is
significant in the x-ray light curve obtained by analysis of the RXTE
All-Sky Monitor (ASM) during the  years 1996-2000. The RXTE/ASM
ephemeris is extended by analysis of all historical data (Vela 5B and
Ginga) to yield a period 171.033$\pm$0.326 days with no evidence for
period change ($|{\dot{P}/ {P}}| < 2.20 \times 10^{-4}$ yr$^{-1}$).
All reported x-ray burst activity is confined to within $\pm$23 d of
the predicted minima.  This stable long-term modulation is consistent
with 4U 1820$-$30 being a hierarchical triple system with a $\sim$1.1
d period companion.

\end{abstract}

\keywords{accretion, accretion disks --- stars individual (4U 1820$-$30) --- x-rays stars}

\section{INTRODUCTION}

The low-mass x-ray binary (LMXB) 4U 1820$-$30 near the center of the
globular cluster NGC6624 (Grindlay et al. 1984) has a binary orbital period of
685.0118 sec (Stella, Priedhorsky \& White 1987; Smale, Mason \& Mukai
1987; Morgan, Remillard \& Garcia 1988; Sansom et al. 1989; Tan et
al. 1991; van der Klis et al. 1993a,b) and was the first
x-ray burster identified with a known x-ray source (Grindlay et al. 1976).  Its short orbital period
and its x-ray burst activity imply it is a system with a 0.06-0.08 $M_{\odot}$ helium white dwarf secondary star accreting mass onto a primary
neutron star (Rappaport et al. 1987). A 176 d period long-term
variation between high-luminosity and low-luminosity by factor of
$\sim$3 was observed by the Vela 5B spacecraft (Priedhorsky \& Terrell
1984, hereafter PT84).  Quasi-periodic oscillations (QPO) with
various frequencies were also reported (Stella, White \&
Priedhorsky 1987; Hasinger \& van der Klis 1989; Smale, Zhang \& White
1997; Zhang et al. 1998; Wijnands, van der Klis \& Rijkhorst 1999;
Kaaret et al. 1999). 

Whereas the x-ray observations show an 11 minutes sinusoidal-like,
small amplitude ($\sim 3\%$ peak to peak) modulation, Anderson et
al. (1997) discovered a large $\sim$ 16\% (peak to peak) modulation
(period $687.6 \pm 2.4$ sec) in the UV band (wavelength in the 126-251
nm range) from the Hubble Space Telescope (HST).  The modulation may come
from the variable thickness of the outer disk rim (Stella, Priedhorsky
\& White1987). The stability of the period ($\dot
P/P =(-5.3 \pm 1.1) \times 10^{-8}$ yr$^{-1}$, van der Klis et al. 1993b)
makes it certain that the 685 sec modulation is the orbital period.  However, the
negative period derivative is inconsistent with the lower limit ($\dot
P/P >+8.8 \times
10^{-8}$ per year) of the standard scenario proposed by Rappaport et al.
(1987).

The 4U 1820$-$30 luminosity variation was first discovered by
Canizares and Neighbours (1975) and its possible $\sim$176 d periodicity was
first reported by PT84.  The fact that x-ray
bursts have only been seen in the low luminosity state (Clark et
al. 1977, Stella, Kahn \& Grindlay 1984)
indicates that the long-term modulation is an intrinsic change in the 
accretion rate rather than an extrinsic absorption. However, significant
phase shifts from the expected minima predicted by the ephemeris of
the 176.4$\pm$1.3 d period proposed by PT84 have been observed from
the EXOSAT (Haberl et al. 1987) and Ginga (Sansom et
al. 1989).  The $\sim$176 d periodicity, if stable, of the luminosity
variation implies (Grindlay 1988) that 4U 1820$-$30 may be a
hierarchical triple in which a third companion star with a period
of $\sim$1.1 d orbits the 11 minute binary and thereby induces  and an inner binary eccentricity
precession (Mazeh and Shaham, 1979) with a period of $\sim$176 d.

In this paper, we describe our RXTE 1996-1997 PCA and 1996-2000 ASM
observations of 4U 1820$-$30 (section 2) and report the timing analysis
of the data (section 3), including the 11 minute binary periodicity,
phase jitter, period stability, updated quadratic ephemeris, and
search for the $\sim$1.1 d period that could be associated with a third companion.  Analysis of RXTE
ASM data gives a new ``176 d'' modulation ephemeris which connects all
the observations from 1969 to 2000 and exhibits no significant period
derivative.  In section 4, we discuss possible models for the observed light
curve behavior of 4U 1820$-$30.

\section{RXTE OBSERVATIONS}

\indent The RXTE PCA/HEXTE pointed observations of 4U 1820$-$30 were made on 1996
October 26, 28 and 30, and at least once per month between 1997 February
and September.  The observation time interval
spanned about two 176 d luminosity cycles.  Details of the 
RXTE PCA/HEXTE observations are listed in Table 2 in Bloser et
al. 2000.  The data used for the analysis are in PCA (PCU 0 and 1) Standard-2 format with a time resolution of 16 sec.  We divide the data into 4 energy bands,
1.72-3.18 keV, 3.18-5.01 keV, 5.01-6.84 keV and 6.84-19.84 keV.  Since
our timing analysis did not suggest any significant energy dependence
of the patterns observed, we will present in this paper only the
analysis results for band 4 (6.84-19.84 keV) unless otherwise specified. A typical
PCA Standard-2 light curve is shown in Figure~\ref{tylc}. A 
complete analysis of the spectra and spectral variations of 4U 1820$-$30 
for these RXTE observations is presented 
by Bloser et al. (2000).

Data for the 4U 1820$-$30 RXTE/ASM x-ray light curve (2-12 keV) analyzed in this
paper were collected from 1996 January 11 to 2000 March 2 (Figure~\ref{asmlc1d}).  The observation window (1576 days) spanned about 9 contiguous 176 d modulation cycles.  In each modulation cycle, the count rate varies from $\sim$5 to $\sim$35 counts/sec.

\section{DATA ANALYSIS}

\subsection{Binary Periodicity and Phase Analysis}

\indent All RXTE PCA data were first corrected for the barycenter arrival
times.  In order to avoid the possible alias from the $\sim 176$ d
period long-term modulation, we removed the DC term from the observed light
curve.  We carried out a $\chi^2$ analysis of the folded light
curves (32 bins per period) to search for the best
period near 685 sec for the entire 4U 1820$-$30  RXTE PCA data. 
The maximum $\chi^2$ (deviation of folded light curve from constant
flux) is at 685.014 sec as shown in
Figure~\ref{fesall}. Fitting the  685.014 sec peak with a Gaussian returns
a best period of 685.0144 $\pm$ 0.0054 sec. The side bands are
primarily artifacts of the alias period from the observation gaps.  

For comparison with historical results, we folded the light curve of
each observation by the ephemeris from Tan et al. (1991).

$${T_{N}}^{max}= HJD2442803.63544 + (685.0118/86400) \times N, \eqno(1)$$ 

\noindent where N is the cycle count.  The phase corresponds to the maximum of the sinusoidal fit of each
 folded light curve.  A typical folded light curve is shown as 
Figure~\ref{foldlc}. Figure~\ref{xteph} shows the phases of the RXTE
 1996-1997 observations.  The phase of maximum flux are scattered around -0.2 with $\sim0.061$
 (rms) phase jitter.

To update the ephemeris, we appended  the mean arrival times from the RXTE observations to the historical results from SAS-3
(Morgen et al. 1988), Ariel V (Smale et al. 1987), Einstein (Morgan
et al. 1988), Tenma (Sansom et al. 1989), EXOSAT\footnote {Arrival
time errors of 0.0002 d were quadratically
added, see van der Klis et al. (1991a)}  (Stella, Priedhorsky and
White 1987), Ginga (Sansom et al. 1989, Tan et al. 1991 and  van der
Klis et al. 1993a) and ROSAT (van der Klis et al. 1993a,b).  The
phase (or arrival time offset) errors, however, need to be re-estimated.  Van
der Klis et al. (1993b) discovered the phase shifts about
0.038 in the three ROSAT observations in 1991 and 1993, and the historical
phase jitter $\sim$0.050 around the best fit linear ephemeris.
For the RXTE data in this paper, we found the phase jitter is
$\sim$0.061 (=0.00048 d).  Therefore, an additional 0.061 phase error was
quadratically added to the historical data.  For
the RXTE data, we weighted-averaged the phases 
 from each observation for the 1996 and 1997 datasets separately.  We
calculated the mean phases errors from quadratically adding the 0.061
phase jitter to the average phase errors from a sinusoidal fit of the
folded light curves.  The resulting mean phases are $-0.20259 \pm
 0.0613$ and $-0.20352 \pm 0.0612$ for the 1996 and 1997 datasets
respectively.  The mean arrival time for the average phases of two datasets were obtained from the expected flux maxima of the mid observation times of the observation windows, that is

$$T_{mean} = T_{0} + P_{fold} \times (N_{mid} + \phi_{ave}) \eqno(2)$$

\noindent where $T_{0}=HJD2442803.63544$, $P_{fold}=(685.01180/86400)$ d,
$N_{mid}$ is the cycle number closest to the mid time of the dataset
and $\phi_{ave}$ is the mean phase for the 1996 and 1997 observations.

The period derivative can be
obtained from a quadratic fit

$$\Phi = \Phi_0 + {{\Delta P} \over {(P_{fold})^2}} t + {1 \over
2}{{\dot P} \over {({P_{fold}})^2}} t^2\eqno(3)$$

\noindent We applied linear ($\dot P$ = 0) and quadratic fits to the data. Both
fits give acceptable results -- $\chi^2$=12.47 (d.o.f.=21) for linear fit
and $\chi^2$=6.96 (d.o.f.=20) for quadratic fit.  However, the F-test (Bevington 1992)
shows that F($\nu_1$=$\Delta \nu$=1; $\nu_2$=20)=15.83 for linear and quadratic fits.  This implies that the quadratic fit
is better than the linear fit at the $\sim$99.99\% confidence
level. Thus, the quadratic ephemeris is still required.  

From the quadratic fit of the data (Figure~\ref{quadeph}), we obtained $\dot P = (-7.54 \pm 3.21) \times
10^{-13}$ sec/sec or $\dot P/P = (-3.47 \pm 1.48) \times
10^{-8}$ yr$^{-1}$, which is consistent with the value found by van der
Kils et al. (1993b).  The
quadratic ephemeris\footnote{The small offset due to leap seconds was ignored in our data analysis.  The offset is
about 16 seconds from zero phase epoch
(JD2442803) to end of observations (September 9 1997=JD2450701).  If we
assume that the offset drifts linearly with time (first order approximation), this systematic
effect will give only 1.6$\times$10$^{-5}$ sec offset in the second term
of eq. (4).  It is much smaller than the error (1.02$\times$10$^{-4}$
sec) from quadratic fit. The periodic systematic effect due to the difference
between the heliocentric time and the barycentric time ($\sim$2.5 sec,
mainly
determined by the position of Jupiter)
was also neglected.} can be written as

$$ T_{N} = HJD(2442803.63564 \pm 2.2 \times 10^{-4}) +(685.0119 \pm
1.02 \times 10^{-4})/86400
\times N $$ $$+ (-2.99 \pm 1.27) \times 10^{-15} \times N^{2}
\eqno(4)$$

Tan et al. (1991) showed that the 685 sec modulation phases are well
fitted by a period of $\sim$8.5$\pm$0.2 yr sinusoidal curve from the 1976-1989
observation results.  The period may be real or an artifact from the
observation gap between 1981 and 1984 (Tan et al. 1991).  To clarify
the ambiguity, we used the constant-sinusoidal, linear-sinusoidal, and
quadratic-sinusoidal models to fit the phases from all the 1976-1997 observations near the 8.5 yr period.  The $\chi^2$ minimum fit results for the three different
models are listed in Table 1.  Although the linear-sinusoidal and the
quadratic-sinusoidal models gave smaller reduced $\chi^{2}$ values
than the quadratic
model, the fitted amplitudes for both cases were only
$\sim$0.05 (modulation period $\sim$6.5 yr), less than the 0.06 phase jitter.  Therefore, there is no
significant long-term phase periodic modulation of period $\sim$6-8
years and the $\sim$6.5 year ``period'' is
highly likely to be an artifact from the phase jitter and the observation gaps between 1976 SAS-3
and 1985 EXOSAT ($\sim$3500 d, 1.5 period), 1985 EXOSAT and 1989 Ginga
($\sim$1300 d, 0.5 period), and the 1989 Ginga and 1991 ROSAT
($\sim$1400 d, 0.5
period) observations.

In explaining the discrepancy between the positive period derivative
predicted by the standard scenario vs. the negative observed $\dot P$, van der Klis et al. (1993b) demonstrated that
the negative phase shift could be due to a long-term variation of the disk size.
The proposal can be tested by looking for a dependence of orbital phase
on $\dot M$ (van der Klis et al. 1993b), or $L_x$.  We compared the 11 minute
modulation phases from the 1996-1997 the RXTE observations and the simultaneous
count rates from the RXTE All Sky Monitor (except February 9 1997
observation which has no All Sky Monitor data).  The linear
correlation coefficient is only 0.17, which implies that the
uncorrelated probability is about 70\%.  Therefore, no significant
correlation between binary orbital phases and luminosities is
observed.  Further considerations about the phase shift are given in section 4.

\subsection{Possible Period Side Bands}

Grindlay (1986, 1988) suggested that the period of $\sim$176 d long-term
modulation of 4U 1820$-$30 (PT84) may be due to a hierarchical triple companion
star (captured by the compact binary in the high
density cluster core) which modulates the eccentricity of the inner binary at a long-term period $P_{long} = K P^{2}_{outer}/P_{inner}$,
where $P_{inner}$ and $P_{outer}$ refer to the binary period and the
orbital period of the third companion and K is a constant of order
unity which depends on mass ratios and relative inclinations (Mazah
and Shaham 1979).  Under the
triple model, with 176 d long-term modulation and 685 sec binary
orbital period, the theoretical orbital period of the third star in
the 4U 1820$-$30 system would be $\sim$1.1 d (for K$\simeq$1, however,
factors of 2 smaller or larger periods for the triple companion could
be accommodated for differing inclination).  If 4U 1820$-$30 is a triple
system, the 685 sec modulations would be affected by such a period and the
beat side bands should appear near the peak of the power spectrum.  We
considered the binary motion around the center of mass of triple
system.  For the third companion star of mass $\sim$0.5M$_\odot$
(approximate maximum allowed by the optical counterpart) and
$\sim$1.1 d orbital period, the radius of the binary motion relative to
the center of mass of triple system is only $\sim$3.4 sin$i_3$
light-second, where $i_3$ is the inclination angle of the orbit of
triple companion.  In other words, the
observed $\sim$1.1 d period phase modulation amplitude is no
more than 5$\times$10$^{-3}$.  Although the phase variation
form the binary motion may be too
small to be observed,  the third companion star could still affect the
light curve in other ways.  If, for example, the third companion star
makes the 11 minute modulation amplitude change,  $\sim$1.1 d beat side
bands\footnote{The beat side band periods $P_{beat}=(1/P_{inner} \pm n/P_{outer})^{-1}$ where n is
positive integer and $P_{inner}$ and $P_{outer}$ are the binary (11 minutes) and third star
period ($\sim$1.1 d), respectively.  The first harmonic (n=1) beat side band periods
are thus 689.97 seconds (f=1.45$\times$10$^{-3}$ Hz) and 680.10
seconds (f=1.47$\times$10$^{-3}$ Hz).  The apparent sideband
peak in the top plot of Figure~\ref{clean} are tantalizing but
probably due to the $\sim$1 d spacing between successive
observations.  A longer continuous observation would be required to
remove theses alias peaks.} amplitudes may be detectable
in the Fourier spectrum.

To search possible side bands, we considered only the 1996 October 26 to 30
light curves  because the observation gaps for the 1997
observations were too significant.  To further minimize the observation windows (from
observation gaps and Earth occultation), a one-dimensional CLEAN
algorithm described by Roberts, Lehar \& Dreher (1987) was applied to
convert for the unequally spaced observations. We searched an
arbitrary wide frequency range between $1.38\times 10^{-3}$ Hz (P=724.6 sec) and $1.55
\times 10^{-3}$ Hz
(P=645.2 sec) with amplitudes greater than 2$\sigma$ significance.  The
search results are shown as Figure~\ref{clean}.  Only one primary peak is observed.  The peak has an amplitude of 2.42 $\pm$ 0.26 cts/sec and a period of 685.120 sec.  There are no other significant side bands beside this primary peak.

\subsection{$\sim$176 Days Modulation}

The 1-day average binned light curve for the RXTE ASM data (see
Figure~\ref{asmlc1d}) shows a clear modulation with relatively
rapid rise and slow fall in count rate with a $\sim$176 d period.  The
count rate varies from 5 cts/sec to 35 cts/sec (see
Figure~\ref{asmlc1d}) throughout each cycle.  ``Inter dips'' were also
observed between 1st-2nd, 4th-5th, 5th-6th and 6th-7th minima of the
RXTE ASM light curve (minimum cycle count indices are
marked on the bottom plot of Figure~\ref{asmlc1d}).

PT84 reported the 4U 1820$-$30 long-term modulation period to be
176.4$\pm$1.3 d.  However, the ephemeris from PT84 ($T_{min} =
JD2442014.5 + (176.4 \pm 1.3)\times N$) does not match the EXOSAT 1985 August 19/20 4U 1820$-$30 low state observation
(Haberl et al. 1985) where the expected minimum (by ephemeris from PT84)
was offset by 48 days.  A similar discrepancy was found in the data
from the Ginga all sky monitor (delayed by $\sim$50 d, Sansom et
al. 1989,  S. Kitamoto, private communications).  The light
curve of the RXTE ASM data also shows a $\sim$0.45 (80d) phase shift (see
top plot of Figure~\ref{asmlc1d}).  The inconsistency may be caused by an incorrect
ephemeris (phase zero epoch, period or both) or a period drift.

To obtain the best period to describe the 4U 1820$-$30 RXTE ASM
light curve, we first applied a fast Fourier transformation (FFT).
However, because the observation window is only a brief $\sim$8.8
cycles (1576
days), the FFT frequency resolution, $\delta f$= 1/(1576d)= 0.233 cycle
per year, yields the period resolution near 176 d of $\delta P \approx P^2
\times \delta f$=19.8 d, which is too coarse for $\sim$176 d period modulation.
Therefore, the interpolated Fourier transformation (Middleditch, Geich and
Kulkarni 1993) was applied to further determine the best
period.  The FFT is only able to show amplitudes of the ``integer''
frequencies ($f_n=n/T$ where n is an integer and T is the total time of
the data) whereas the interpolated Fourier may show the amplitude of
``non-integer'' frequencies ($f_r=r/T$, where r is any real number).
The non-integer amplitude $A_r$ can be estimated from the locally
neighboring Fourier amplitudes as

$$ A_{r} \sim \sum_{l=[r]-m}^{[r]+m} A_{l} e^{-i\pi(r-l)}
{{sin(\pi(r-l))}\over {\pi(r-l)}} \eqno(5)$$

\noindent where m is integer and [r] denotes the nearest integer of
r.   The uncertainty of peak frequency

$$ \sigma_{f} = {{3} \over {\pi \alpha T \sqrt{6P_0}}} \eqno(6)$$

\noindent where

$$ \alpha = {1 \over {\pi T}} \sqrt{- {3 \over {2P_0}} {{\partial^2P}
\over {\partial f^2}}} \eqno(7)$$

\noindent $P_0$ the peak power, and T is the length of the observation
window (Middleditch, Geich and Kulkarni 1993).

We chose m=2 and the resolution of r to be 0.1.  The interpolated
Fourier transformation spectrum is shown in the bottom plot of
Figure~\ref{asmfft}.  The peak amplitude was observed at f=2.130
cycle/yr with a value of 4.506 cts/sec. The frequency uncertainty from
eq. (6) and eq. (7) is 0.0240
cycle/yr where the second derivative in eq. (7) is estimated by the
quadratic fit around the peak of the power spectrum.  Therefore, the best 4U 1820$-$30
long-term modulation period for the RXTE ASM data is
$171.39\pm1.93$ d.  Furthermore, the $\chi^2$ period searching method
gave the best period of 171.23 d ($\pm$7.36 d), close to the interpolated
Fourier transformation result.

The RXTE ASM light curve, derived by folding at a period of 171.39 d
(see Figure~\ref{asmfoldlc}), shows
that the minimum closest to the mid of observation is at JD2450907.96
and $\pm$0.07 (rms) phase jitter (or $\pm$12 d).  The best linear ephemeris to
describe the intensity minimum of the RXTE ASM light curve (hereafter
local ephemeris) can be written as

$$ T_{min}^{RXTE} =  JD2450907.96 \pm 12.00+ (171.39 \pm 1.93) \times N. \eqno (8)$$

To obtain the best linear ephemeris for the intensity minimum in the RXTE
ASM light curve and historical data, we assigned an uncertainty
$\pm$12 d (from 0.07 phase jitter obtained from the RXTE ASM light curve) to the
minimum time reported by PT84 (JD2442014.5).  Combined with intensity
minimum times observed by the Ginga all-sky monitor (JD2446822$\pm$26,
S. Kitamoto, private communications) and RXTE ASM data, the linear fit
yields the best historical ephemeris of

$$ T_{min} =  JD2450909.90 \pm 11.66 + (171.033 \pm 0.326) \times N. \eqno (9)$$

\noindent The expected intensity minimum times from eq. (9) vs. the RXTE ASM light
curve are shown as the bottom plot of Figure~\ref{asmlc1d}.  An
independent check on the ephemeris may be derived from timing of x-ray
burst activity. Stella, Kahn and Grindlay (1984 and references therein)
reported that no bursts are detected in the high state, which implies that
the ``burst phase'' should be near phase zero.  Table
2 lists the observation dates with bursts being detected and the
phases of these days. No burst phase folded by eq. (9) exceeded the
range $\pm$0.13
($\pm$23 d).  This indicates that the long-term modulation period
of 4U 1820$-$30 is close to 171.033 d and  stable over $\sim$30
years.

To estimate the period derivative (or its upper limit), we suppose the
minimum times ($T_N$) obey a quadratic ephemeris (for small $\dot P$)

$$ T_N =  T_0 + P_0 N + {1 \over 2} P_0 \dot P N^2, \eqno (10)$$

\noindent where N is cycle count index.  Taking the phase zero epoch
$T_0$ to be that in eq. (8) (JD2450907.96), since the cycle
numbers are small for RXTE intensity minima (from -4 to 4), the
quadratic term in eq. (10) can be neglected.  Eq. (10) is reduced to a
linear ephemeris equal to the local ephemeris (i.e. eq. (8)).  The
phases of historical data (Vela 5B and Ginga) folded by eq. (8) should
be due to the period derivative

$$\Phi =   {1 \over 2}{{\dot P} \over {({P_0})^2}} \Delta
t^2, \eqno (11)$$

\noindent where $\Delta t$ is time difference between minimum time and
$T_0$.  Applying eq. (8) to historical data, we found no significant
period derivative.  The 2$\sigma$ (90\%) 
confidence level upper limit for the change in 171 d period is  
$|\dot P| < 1.03 \times 10^{-4}$ days per day (= 0.038 days per
year) or $|{\dot{P}/{P}}| < 2.20 \times 10^{-4}$ yr$^{-1}$.  This
stable long-term modulation is
consistent with 4U 1820$-$30 being a hierarchical triple system with a
$\sim$1.1 d period companion.

\section{DISCUSSION}

The observed negative period derivative from the 4U 1820$-$30 RXTE 1996-1997
observations combined with all the historical data is consistent with the previous conclusions
proposed by Tan et al. (1991) and van der Klis et al. (1993a,b).  The
decreasing 685 sec period deviates from the positive $\dot P$
($\dot P/P > 8.8 \times 10^{-8}$ yr$^{-1}$) predicted by the standard scenario 
(Rappaport et al. 1987).  Tan et al. (1991) suggested that the
discrepancy is probably caused by the acceleration of the binary star by
the gravitational potential of the globular cluster.  On the other hand,
by analyzing the theoretical minimum of the period derivative, van der
Klis et al. (1993a) found that the gravitational acceleration by the
globular cluster is not enough to explain the observed results even if
the line of sight is very close to the line connecting the binary with the
center of the cluster at the projected separation 4$\pm$1 arcsec.

However, King et al. (1993) measured the NGC6624 cluster center with
the HST Faint Object Camera (FOC) and discovered that 4U 1820$-$30 is
0.66 arcsec from the cluster center.  For an assumed distance of
6.4 kpc (Vacca et al. 1986, Haberl and Titarchuk 1995), this is equivalent to only 0.02 pc (projected)
 from the core.  Using the model proposed
by van der Klis et al. (1993a), the maximum gravitational acceleration
along the line of sight could be a/c=$2.5 \times 10^{-15}$ sec$^{-1}$.
Combining the period derivative derived in section 3.1 and its value
from the standard scenario, we determined the acceleration along the line of
sight to be a/c=$3.9 \times 10^{-15}$ sec$^{-1}$, only $\sim$50\% larger
than the maximum value.  Therefore, given the uncertainties both in $\dot
P$ and the cluster acceleration (i.e. center and mass model), gravitational acceleration by the
globular cluster is still a possible explanation for the negative
period derivative (also see King et al. 1993).

Another potential explanation of the negative period derivative (or,
negative phase shift) was proposed by van der Klis et al. (1993b).  The 
685 sec intensity modulation of the 4U 1820$-$30, as for the dipping sources, is
probably caused by the occultation by the accretion stream of the vertical structure at the edge of 
the accretion disk (Stella, Priedhorsky and White 1987; Morgan et al. 
1988; Sansom et al. 1989; van der Klis et al. 1993b).  The azimuth point of
impact depends on the disk size.  The bulge on the disk rim
could shift by as much as $\sim -120^{\circ}$, larger than the observed phase
shifts of $\sim -72^{\circ}$ ($-$0.2 phase; the approximate value needed
to account for the negative $\dot P$).  However, if we consider
only the standard scenario, $\dot P/P \sim 8.8 \times 10^{-8}$ yr$^{-1}$, the
phase shift due to the period change is expected to be $\sim +0.75$ from the 1976 SAS3
to the 1997 RXTE observations.  This result implies that the total bulge
phase shift is $\sim -0.95 (-350^{\circ})$.  The bulge is unlikely to
have such a large phase shift.  Furthermore, the disk
size should be highly correlated with the accretion rate and, of course,
the luminosity $L_x$.  As discussed in section 3.1, no significant correlation between orbital phase
and luminosity (171 d variation) is found in the RXTE 1996-1997
data.  We hence conclude that the negative phase shift is unlikely to
be caused by a variation of the disk size.

In this paper, we derived the long-term modulation period and showed
from the ephemeris (eq. (9)) that the 4U 1820$-$30 bursts are only
observed in the low state.  By re-analyzing the historical data as
well as tabulated burst activity time, we derived a period
P=171.033$\pm$0.326 d and a limit on $|\dot P| < 1.03 \times 10^{-4}$
days per day.  The high correlation between the burst activity and
the luminosity suggests that the 171 d modulation is stable and indeed
an intrinsic luminosity change rather than an extrinsic periodic
obscuration.  This luminosity modulation and (primarily) its long-term
stability supports earlier suggestions for a hierarchical triple
companion.  The mass transfer rate is very sensitive to the Roche lobe
radius, which is proportional to the inner binary separation.  A
hierarchical companion third star will induce an eccentricity
variation in inner (11 minutes) binary with period $P_{long}=K
P^2_{outer}/P_{inner}$ (Mazeh and Shaham, 1979).  When the
minimum separation of the inner binary is small, the mass
transfer rate and luminosity changes are enhanced.   The triple model
to 4U 1820$-$30 system implies a $\sim$1.1 d period third companion is
responsible for the 171.033 d long-term intensity modulation.  The
triple companion star affects the orbital motion of the inner binary
through beats of 685 sec binary period and $\sim$1.1 d companion star
orbital period.  Our RXTE observations were not sensitive to this
because of both data sampling ($\sim$1 d observation gaps) and the
small amplitude expected.  An additional test of the triple model
could be conducted by a continuous, or optimally sampled long (\ga
3-10 d) observation of 4U 1820$-$30 to measure the small
($\sim$3.4 sin$i_3$ light-second, see section 3.2) phase shifts, or the possible
modulation side bands (Figure~\ref{clean}) without 1 d sampling alias effects.

\acknowledgments
The authors thank S. Kitamoto for invaluable assistant with the Ginga
data, RXTE-GOF for help with RXTE data analysis, and the HEASARC for
archival RXTE ASM data.  This work was supported in part by NASA grant
NAG5-3293 and NAG5-7393.

\clearpage


\figcaption[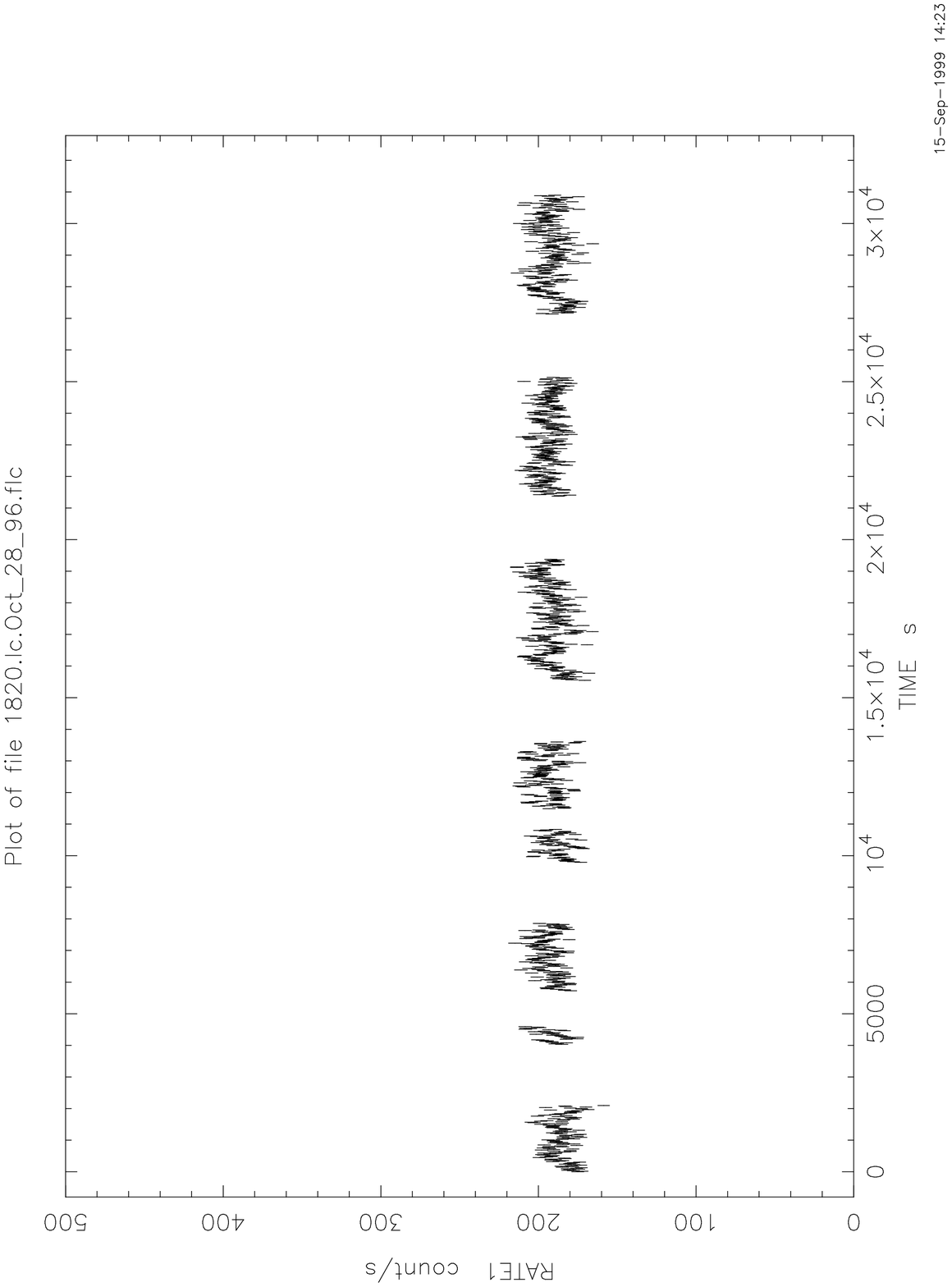]{Light curve of 4U 1820$-$30
(6.84-19.84 keV) observed by RXTE PCA on 1996 October 28/29.  
\label{tylc}}

\figcaption[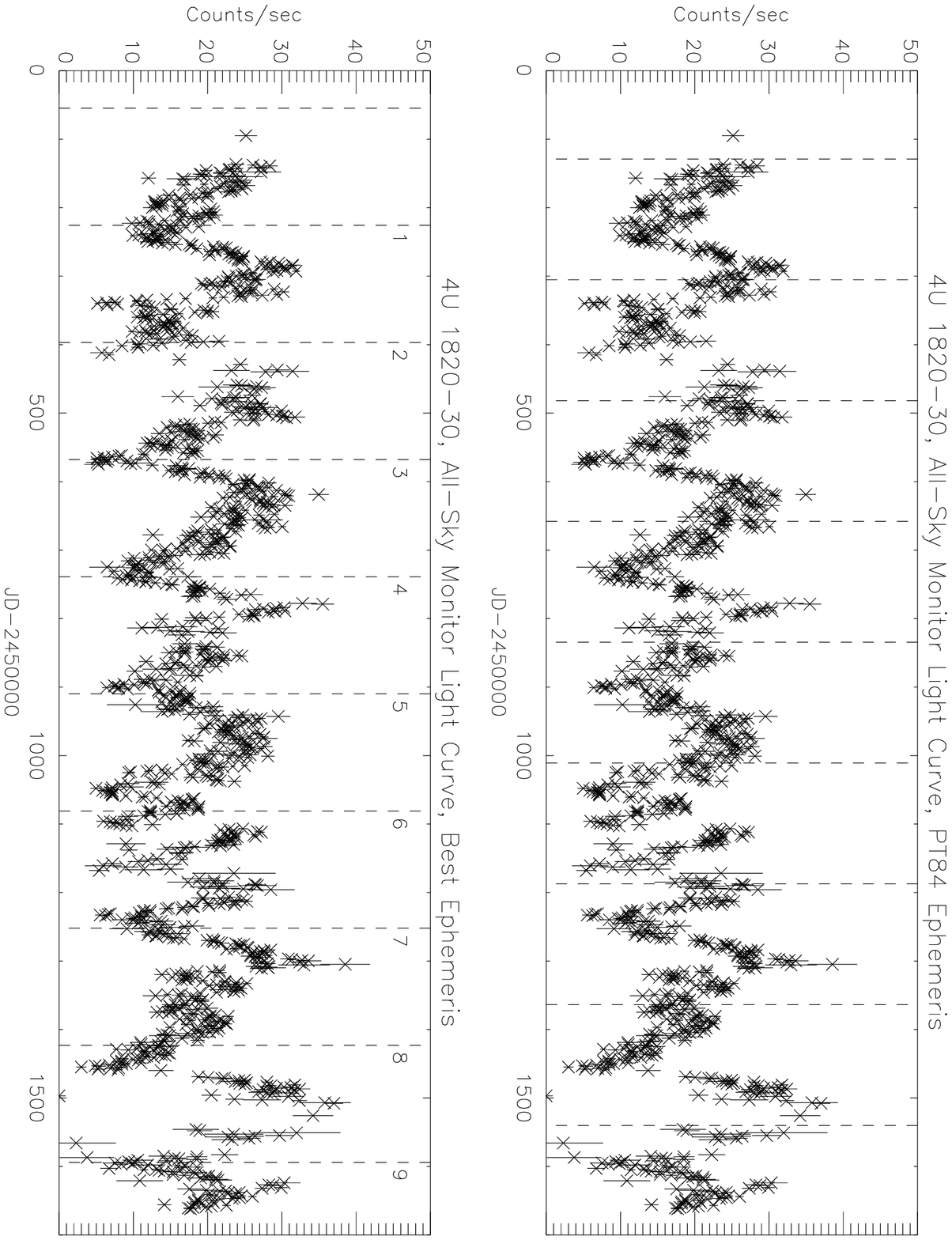]{ The 4U 1820$-$30 light curve observed by
RXTE All-Sky Monitor from 1996 January 11 to 2000 March 2. The dashed
lines are the expected minimum intensity times for the ephemeris from
PT84 (top) and  
ephemeris of eq. (9) in section 3.3 (bottom).
\label{asmlc1d}}

\figcaption[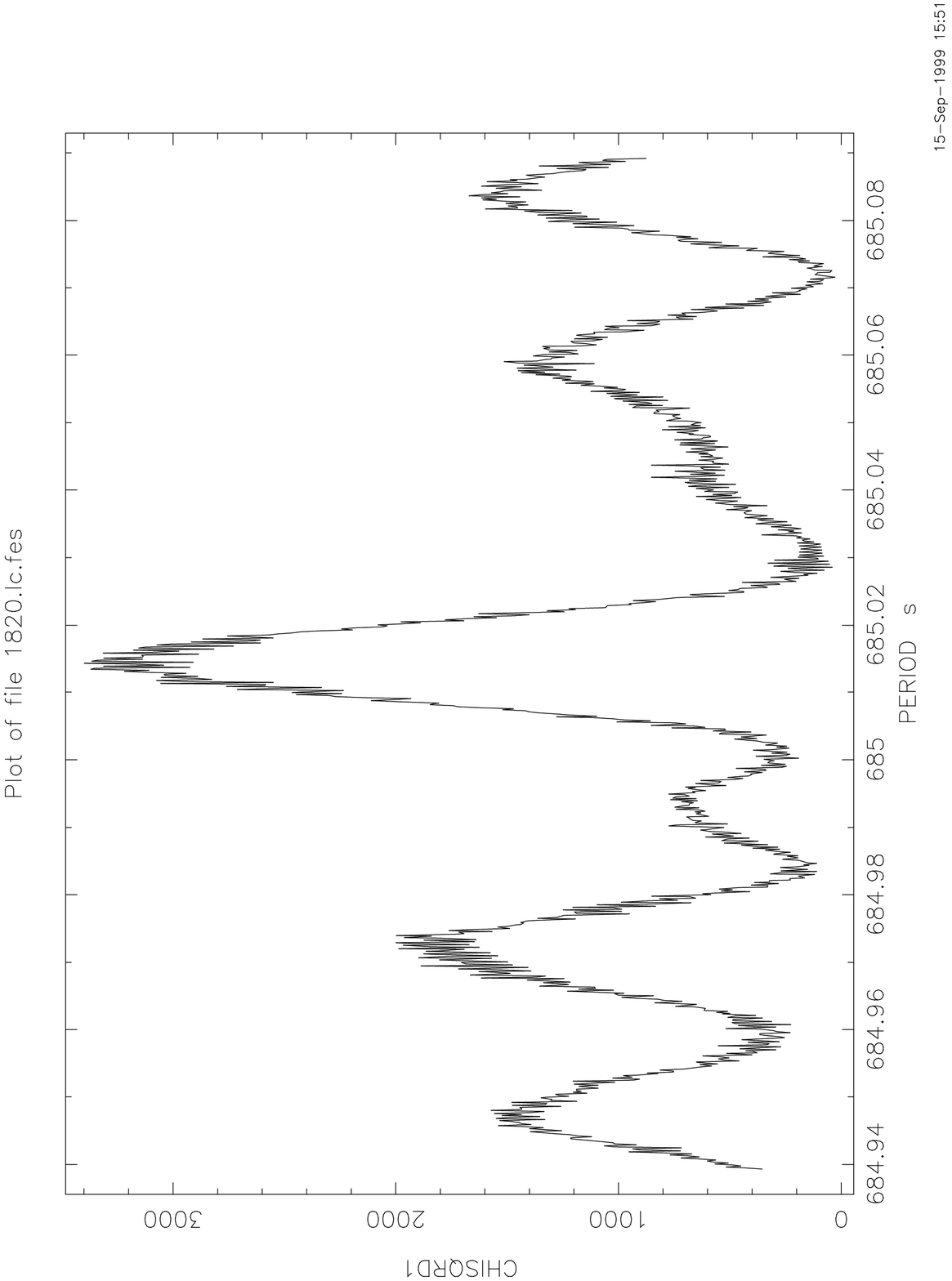]{The result of $\chi^2$ folded period
searching of RXTE PCA 1996-1997 observation.  The Gaussian fit of the peak near 685.01 sec yields a best period of 685.0144 $\pm$ 0.0054 sec.
\label{fesall}}

\figcaption[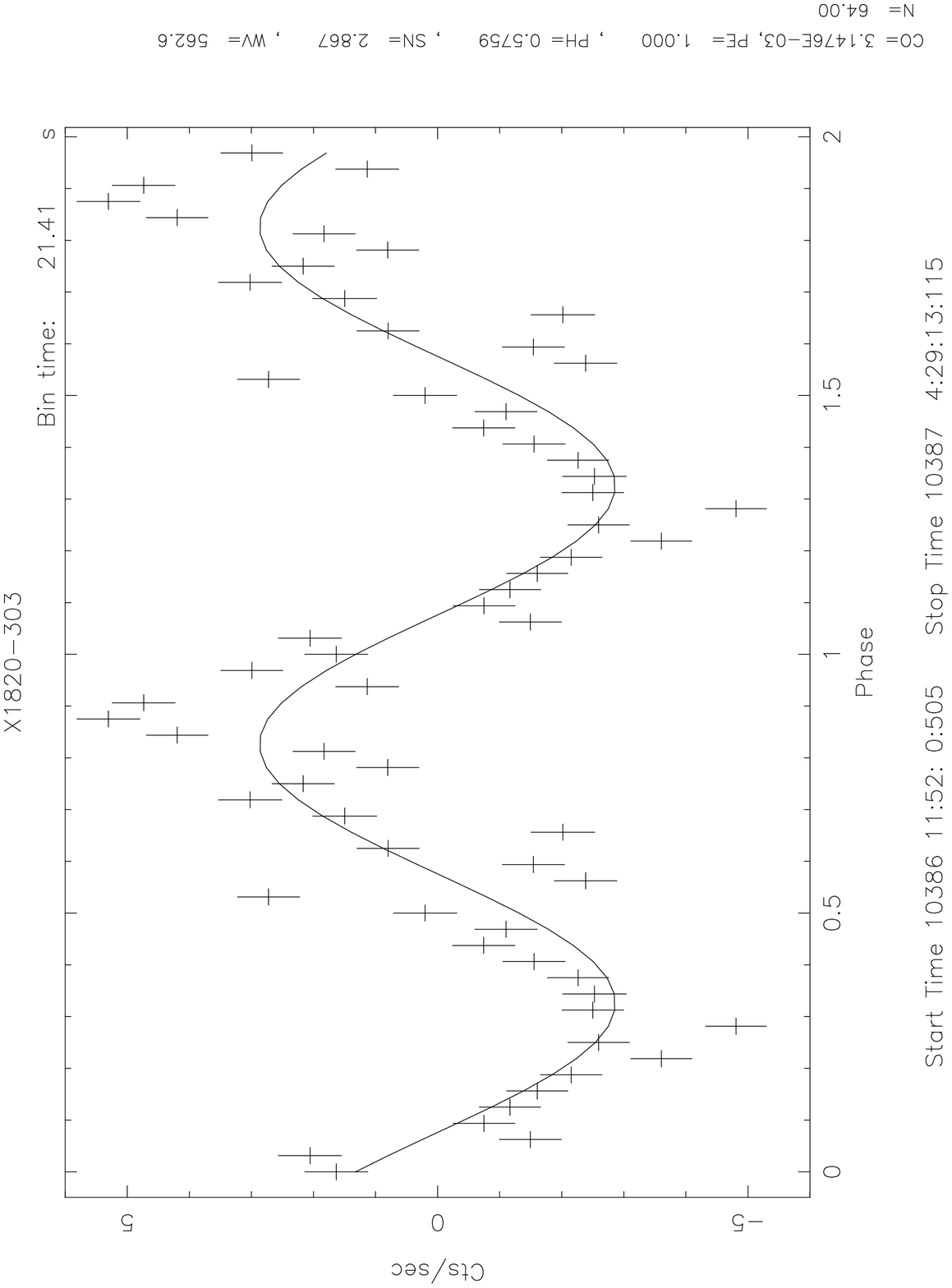]{The folded light curve of the observation
on 1997 September 10 folded by eq. (1) and with the dc flux subtracted.  The maximum of sinusoid fit is at
about phase 0.8 (or -0.2).  Inter-dip is observed at phase 0.5 to 0.6.
\label{foldlc}}
\figcaption[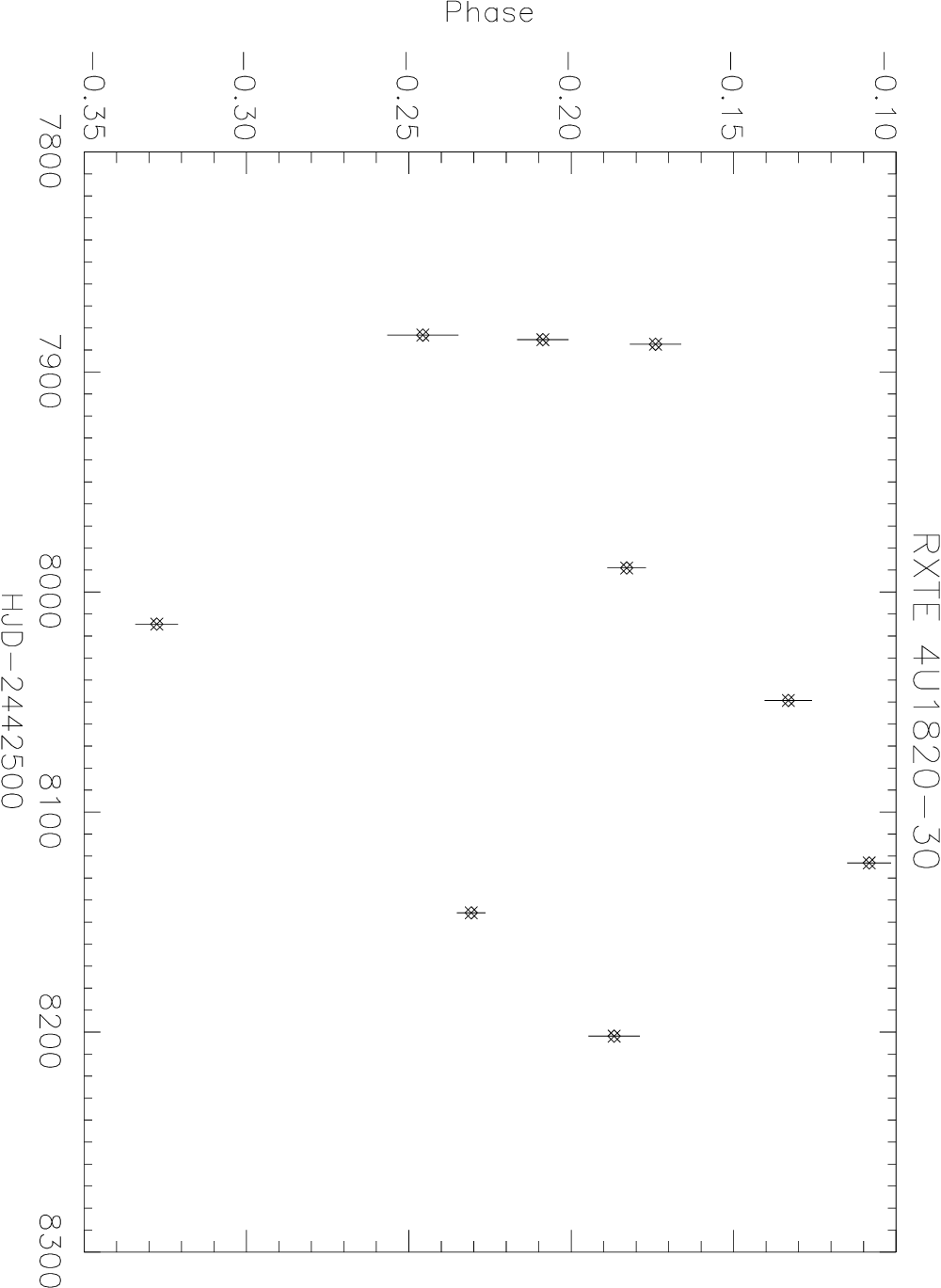]{The phases of 4U 1820$-$30 RXTE
PCA 1996-1997 observation.  The mean fluctuation of the phases is
about 0.061.
\label{xteph}}
\figcaption[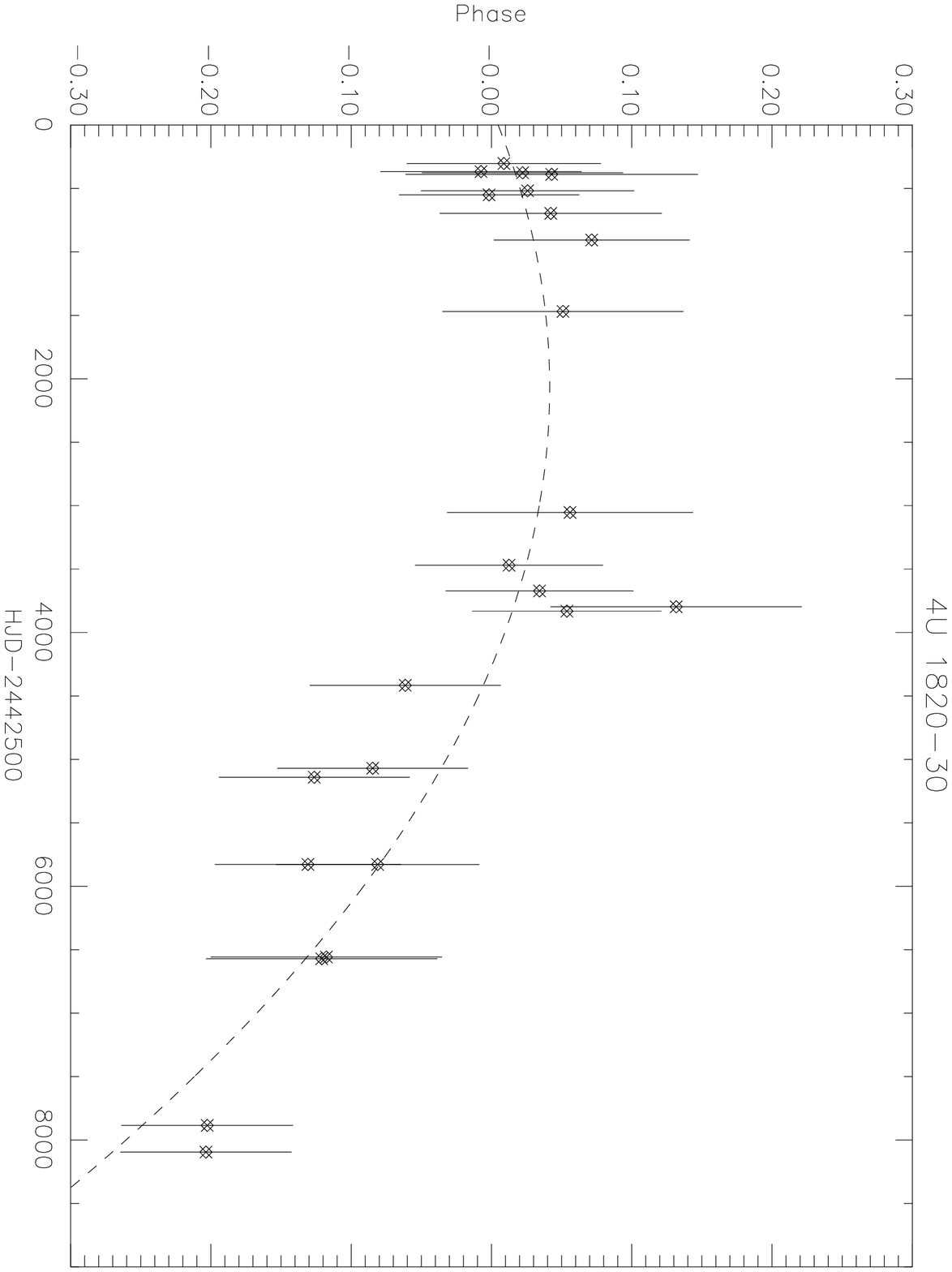]{The phases of 4U 1820$-$30 11
minutes modulation from 1976-1997 observation folded by the ephemeris
of eq (1).  The dashed line represents the best quadratic fit result.
\label{quadeph}}
\figcaption[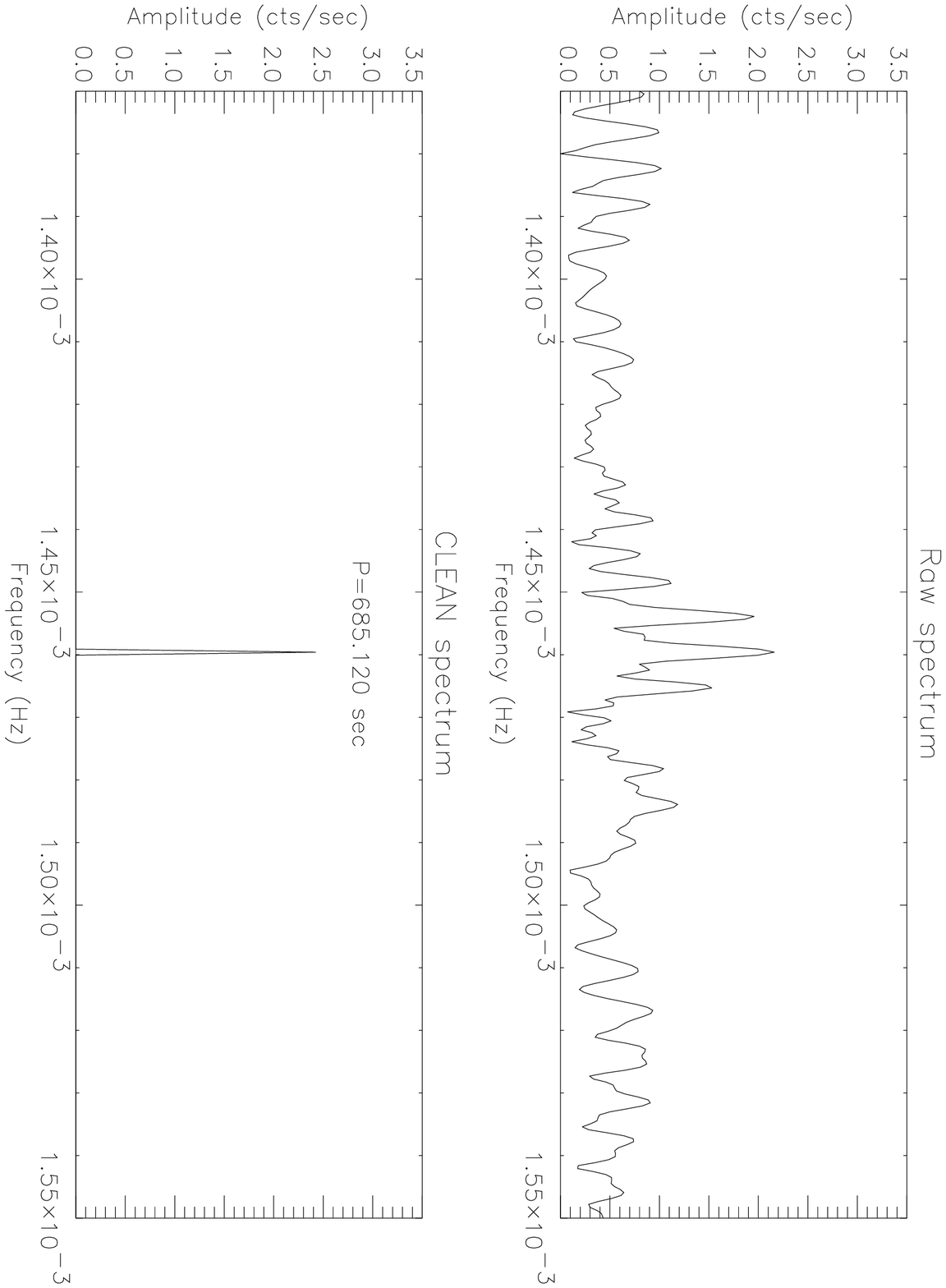]{The raw (top) and CLEAN
(bottom) Fourier amplitude (i.e. absolute values of the Fourier
transformation) spectra of RXTE 1996 October observations.  The
upper amplitude limit for the CLEAN spectrum is 0.52 cts/sec (2$\sigma$).  There is no clear side band beside the peak; therefore, the side bands
in the raw spectrum are likely artifacts from $\sim$1 d observation gaps.
\label{clean}}

\figcaption[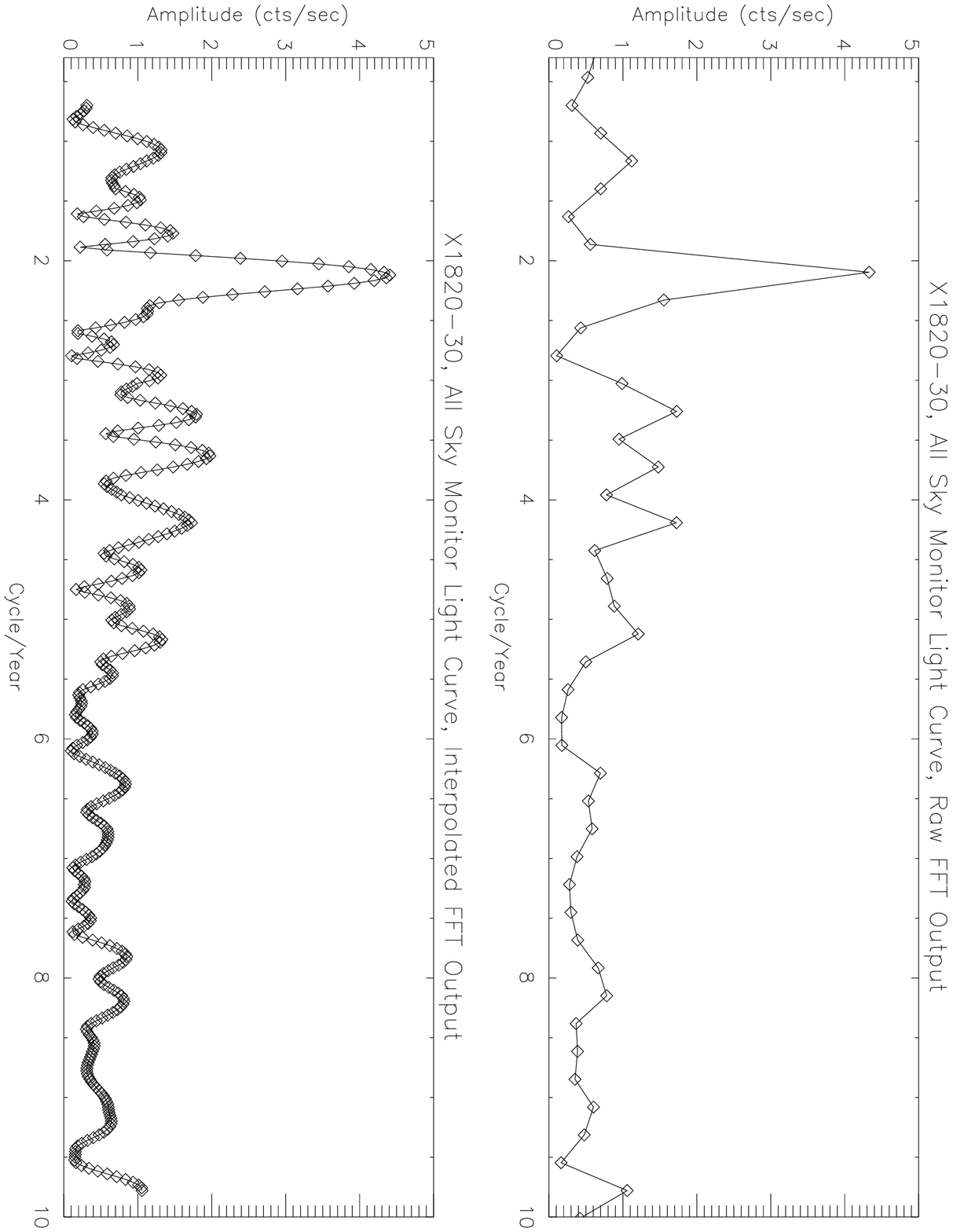]{Raw FFT spectrum (top) and interpolated Fourier spectrum (bottom) of the 4U 1820 RXTE ASM light curves.
\label{asmfft}}

\figcaption[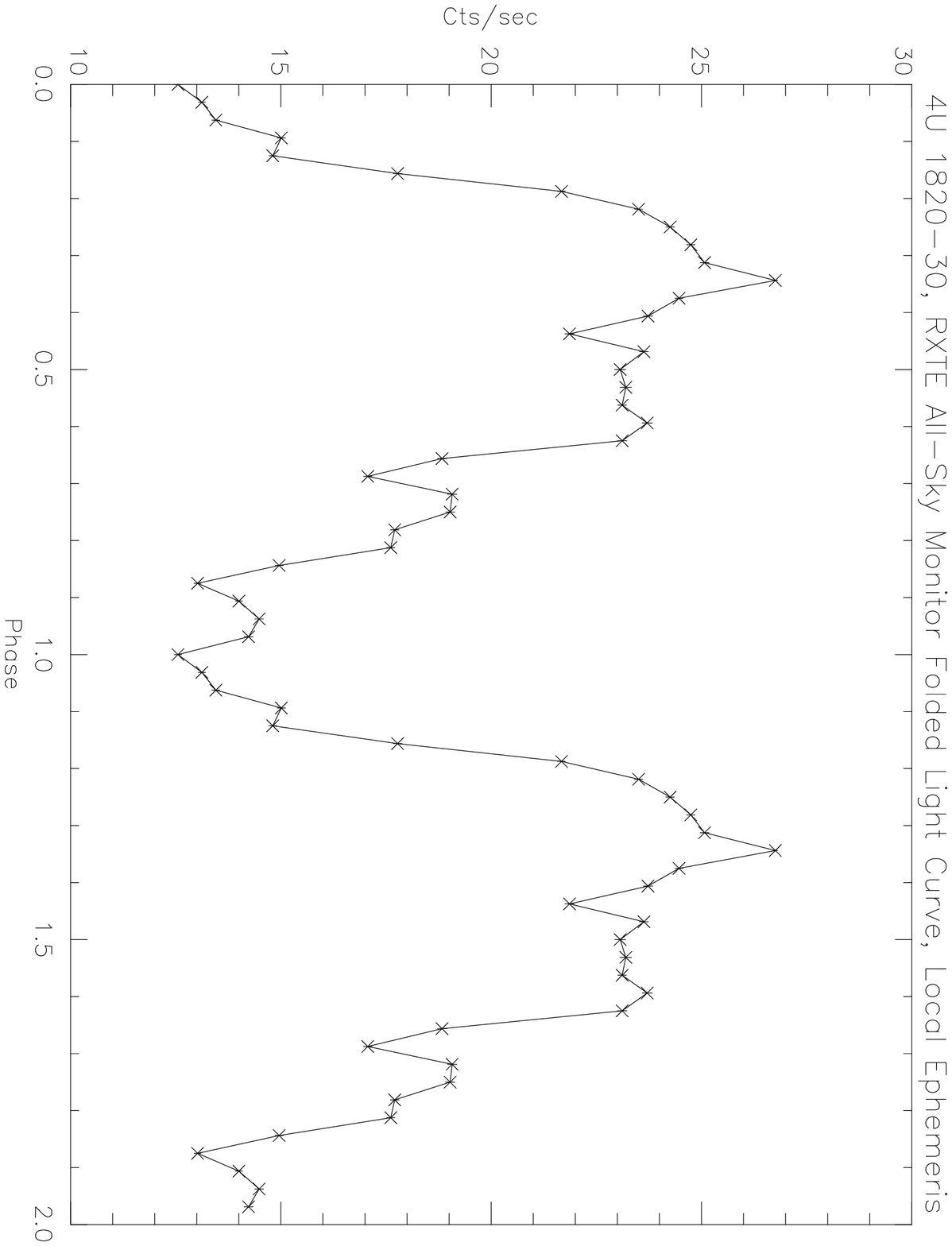]{The 4U 1820$-$30 RXTE ASM light curve
folded by the local ephemeris (eq. (8)).
\label{asmfoldlc}}


\clearpage
\begin{deluxetable}{lccc}
\tablecaption{Phase modulation fit results}
\footnotesize
\tablewidth{0pt}
\tablehead{
\colhead{Model} & 
\colhead{Reduced $\chi^2$} &
\colhead{Period from}&
\colhead{Amplitude from} \\
\colhead{} &
\colhead{} &
\colhead{sinusoidal fit (day)} &
\colhead{sinusoidal fit}
}
\startdata
 L   \tablenotemark{a}&  0.594     &    -  &     -    \\
 Q   \tablenotemark{b}&   0.349    &    -  &     -    \\
 C+S \tablenotemark{c}&  0.782     &  2978.3 $\pm$140.7  & 0.0965 \\
 L+S \tablenotemark{d}&  0.315     &  2475.9 $\pm$224.8 & 0.0567 \\
 Q+S \tablenotemark{e}&  0.223     &  2368.9 $\pm$267.0 & 0.0446 \\

\enddata
\tablenotetext{a}{Linear model}
\tablenotetext{b}{Quadratic model}
\tablenotetext{c}{Constant + sinusoidal model around P=8.5 year}
\tablenotetext{d}{Linear + sinusoidal model around P=8.5 year}
\tablenotetext{e}{Quadratic + sinusoidal model around P=8.5 year}
\end{deluxetable}
\clearpage

\begin{deluxetable}{lcccc}
\tablecaption{X-ray Burst phases}
\footnotesize
\tablewidth{0pt}
\tablehead{
\colhead{Observation date} & 
\colhead{Observation date} &
\colhead{Phase folded by}&
\colhead{Phase folded by}&
\colhead{Reference}\\
\colhead{(dd/mm/yy)} &
\colhead{(JD)} &
\colhead{PT ephemeris} &
\colhead{eq. (9)} &
\colhead{}
}
\startdata
18/05/75  &  2442550.5     &    0.039 &     0.124   & 1   \\ 
28/09/75  &  2442683.5     &    -0.207  &   -0.098   & 2\\
15/03/76\tablenotemark{a}  &  2442852.5    &    -0.250   &   -0.110   & 3 \\ 
20/08/85  &  2446297.5    &    0.280   &    0.032   & 4\\
\enddata
\tablenotetext{a}{Mid observation date of 1976 March 11.5 to 19.5}
\tablerefs{(1) Clark et al. 1976; (2) Grindlay et al. 1976; (3) Clark et al. 1977; (4) Haberl et al. 1987}
\end{deluxetable}

\clearpage

\begin{figure}
\figurenum{\ref{tylc}}
\plotone{f1.eps}
\caption{}
\end{figure}

\begin{figure}
\figurenum{\ref{asmlc1d}}
\mbox{{\resizebox{160mm}{220mm}{{{\rotatebox{90}{\rotatebox{90}{\includegraphics{f2.eps}}}}}}}}
\caption{}
\end{figure}

\begin{figure}
\figurenum{\ref{fesall}}
\plotone{f3.eps}
\caption{}
\end{figure}

\begin{figure}
\figurenum{\ref{foldlc}}
\plotone{f4.eps}
\caption{}
\end{figure}

\begin{figure}
\figurenum{\ref{xteph}}
\mbox{{\resizebox{160mm}{220mm}{{{\rotatebox{90}{\rotatebox{90}{\includegraphics{f5.eps}}}}}}}}
\caption{}
\end{figure}

\begin{figure}
\figurenum{\ref{quadeph}}
\mbox{{\resizebox{160mm}{220mm}{{{\rotatebox{90}{\rotatebox{90}{\includegraphics{f6.eps}}}}}}}}
\caption{}
\end{figure}

\begin{figure}
\figurenum{\ref{clean}}
\mbox{{\resizebox{160mm}{220mm}{{{\rotatebox{90}{\rotatebox{90}{\includegraphics{f7.eps}}}}}}}}
\caption{}
\end{figure}

\begin{figure}
\figurenum{\ref{asmfft}}
\mbox{{\resizebox{160mm}{220mm}{{{\rotatebox{90}{\rotatebox{90}{\includegraphics{f8.eps}}}}}}}}
\caption{}
\end{figure}

\begin{figure}
\figurenum{\ref{asmfoldlc}}
\mbox{{\resizebox{160mm}{220mm}{{{\rotatebox{90}{\rotatebox{90}{\includegraphics{f9.eps}}}}}}}}
\caption{}
\end{figure}


\begin{references}
\reference{and97} Anderson, S. F., et al. 1997, ApJ, 482, L69
\reference{bev92} Bevington, P. R. 1992, Data Reduction and Error Analysis for
the Physical Sciences (New York: McGraw-Hill)
\reference{blo00} Bloser, P. F., Grindlay, J. E., Kaaret, P., Zhang,
W., Smale, A.P., Barret, D. 2000, ApJ, 542, 1000
\reference{can75} Canizares, C. R. \& Neighbours, J. E. 1975, ApJ, 199, L97
\reference{cla76} Clark, G. W., et al. 1976, ApJ, 207, L105
\reference{cla77} Clark, G. W., et al. 1977, MNRAS, 179, 651 
\reference{josh76} Grindlay J. E., et al. 1976, ApJ, 205, L127
\reference{josh84} Grindlay J. E., et al. 1984, ApJ, 282, L13
\reference{josh86} Grindlay J. E. 1986, in The Evolution of Galactic x-ray
Binaries, J. Trumper, W. Lewin and W. Brinkman, eds., Nato ASI Series,
Vol. 167, p. 25
\reference{josh88} Grindlay, J. E. 1988, in IAU Symposium, Globular Cluster
System in Galaxies, ed. J. E. Grindlay and A. G. Davis Philip
(Dordrecht: Reidel), p. 347
\reference{hab87} Haberl, F., et al. 1987, ApJ, 314, L266
\reference{hab95} Haberl, F. and Titarchuk, L. 1995, A\&A, 299, 414
\reference{has89} Hasinger, G. \& van der Klis, M. 1989, A\&A, 225, 79
\reference{kaa99} Kaaret, P., et al. 1999, ApJ, 520, L37
\reference{kin93} King, I. R., et al. 1993, ApJ, 413, L117
\reference{maz97} Mazeh, T. \& Shaham, J. 1979 A\&A, 77, 145
\reference{mid93} Middleditch, J., Deich, W. \& Kulkarni, S. 1993, in
Isolated Pulsar (Cambridge University press), Van Riper, Epstein and
Ho (eds), p. 372
\reference{mor88} Morgan, E. H. Remillard, R. A. \& Garcia, M. R. 1988, ApJ, 324, 851
\reference{pri84} Priedhorsky, W. \& Terrell, J. 1984, ApJ, 284, L17
\reference{rap87} Rappaport, S., et al. 1987, ApJ, 322, 84
\reference{rob87} Roberts, D. H., Lehar, J. \& Dreher, W. 1987, Astron. J., 93, 968
\reference{san89} Sansom, A. E., et al. 1989, PASJ, 41,591

\reference{sma87} Smale, A. P.,Mason, K. O. \& Mukai, K. 1987, MNRAS,
225, 7
\reference{sma97} Smale, A. P., Zhang, W. \& White, N. E. 1997,ApJ, 483, L119
\reference{ste84} Stella, L. Kahn, S. M., and Grindlay J. E. 1984,
ApJ, 282, 713
\reference{ste87a} Stella, L., Peirdhorsky, W. \& White, N. E. 1987, ApJ, 312, L17
\reference{ste87b} Stella, L., White, N. E. \& Peirdhorsky, W. 1987, ApJ, 315, L49
\reference{tan91} Tan, J., et al. 1991, ApJ, 374, 291
\reference{vac86} Vacca, W.D., Lewin, W. H. G., \& van Paradijs, J. 1986, MNRAS, 220 339
\reference{van93a} van der Klis, M., et al. 1993a, MNRAS, 260, 686
\reference{van93b} van der Klis, M., et al. 1993b, A\&A, 279, L21
\reference{wij99} Wijnands, R., van der Klis, M. \& Rijkhorst, E.
1999, ApJ, 512, L39  
\reference{zha99} Zhang, W., et al. 1998 ApJ, 500, L171
\end{references}
\end{document}